\documentclass[]{spie}  
 
\usepackage{amsmath,amsfonts,amssymb}
\usepackage{graphicx}
\usepackage[colorlinks=true, allcolors=blue]{hyperref}
\usepackage{float}

\title{Development and testing of integrated readout electronics for next generation SiSeRO (Single electron Sensitive Read Out) devices
}

\author[a]{Tanmoy Chattopadhyay}
\author[a,b]{Haley R. Stueber}
\author[a,b]{Abigail Y. Pan}
\author[a]{Sven Herrmann}
\author[a]{Peter Orel}
\author[c]{Kevan Donlon}
\author[a,b,d]{Steven W. Allen}
\author[e]{Marshall W. Bautz}
\author[c]{Michael Cooper}
\author[e]{Catherine E. Grant}
\author[e]{Beverly LaMarr}
\author[c]{Christopher Leitz}
\author[e]{Andrew Malonis}
\author[e]{Eric D. Miller}
\author[a]{R. Glenn Morris}
\author[e]{Gregory Prigozhin}
\author[c]{Ilya Prigozhin}
\author[a]{Artem Poliszczuk}
\author[c]{Keith Warner}
\author[a]{Daniel R. Wilkins}

\affil[a]{Kavli Institute of Astrophysics and Cosmology, Stanford University, 452 Lomita Mall, Stanford, CA 94305, USA}
\affil[b]{Department of Physics, Stanford University, 382 Via Pueblo Mall, Stanford CA 94305, USA}
\affil[c]{MIT Lincoln Laboratory, Lexington, MA, USA}
\affil[d]{SLAC National Accelerator Laboratory, 2575 Sand Hill Road, Menlo Park, CA 94025, USA}
\affil[e]{Kavli Institute for Astrophysics and Space Research, Massachusetts Institute of Technology, Cambridge, MA, USA}

\authorinfo{Further author information: Send correspondence to T. Chattopadhyay\\T. Chattopadhyay: E-mail: tanmoyc@stanford.edu}

\begin{document} 
\maketitle


\begin{abstract}
The first generation of Single electron Sensitive Read Out (SiSeRO) amplifiers, employed as on-chip charge detectors for  charge-coupled devices (CCDs) have demonstrated  excellent noise and spectral performance: a responsivity of around 800 pA per electron, an equivalent noise charge (ENC) of 3.2 electrons root mean square (RMS), and a full width half maximum (FWHM) energy resolution of 130 eV at 5.9 keV for a readout speed of 625 Kpixel/s. Repetitive Non-Destructive Readout (RNDR) has also been demonstrated with these devices, achieving an improved ENC performance of 0.36 electrons RMS after 200 RNDR cycles. In order to mature this technology further, Stanford University, in collaboration with MIT Kavli Institute and MIT Lincoln Laboratory, are developing new SiSeRO detectors with improved geometries that should enable greater responsivity and improved noise performance. These include CCD devices employing arrays of SiSeRO amplifiers to optimize high speed, low noise  RNDR readout and a proof-of-concept SiSeRO active pixel sensor (APS). To read out these devices, our team has developed a compact, 8-channel, fast, low noise, low power application specific integrated circuit (ASIC) denoted the Multi-Channel Readout Chip (MCRC) that includes an experimental drain current readout mode intended for SiSeRO devices. In this paper, we present results from the first tests of SiSeRO CCD devices operating with MCRC readout, and our designs for next generation SiSeRO devices.  
  
\end{abstract}




\section{INTRODUCTION}
\label{sec:intro}   
The Single electron Sensitive Read Out (SiSeRO\cite{chattopadhyay22_sisero}), developed at MIT Lincoln Laboratory (MIT-LL), is a novel charge detection technology for X-ray charge-coupled devices (CCDs\cite{Lesser15_ccd,gruner02_ccd,ccd_janesick01}). The SiSeRO design is motivated by floating gate amplifier based devices with high responsivity and sub-electron noise\cite{matsunaga91} and is similar to the design of depletion field effect transistor devices (DEPFETs\cite{kemmer87_depfet,strueder00_depfet_imager}). In SiSeROs, an internal gate replaces the doped floating diffusion implant (FD) employed in source follower CCDs \cite{bautz18,bautz19,bautz20,chattopadhyay22_ccd,Bautzetal2022,bautz24_ccd_spie}, minimizing the parasitic capacitance, with the goal of improving the gain and, therefore, noise performance at high readout speeds. SiSeROs also allow multiple independent measurements of the same pixel charge using Repetitive Non-Destructive Readout (RNDR). This improves the readnoise by a $\sqrt{N}$ factor, potentially into the sub-electron regime \cite{treberspurg22_rndrdepfet}.     

In our first SiSeRO test devices with a single SiSeRO amplifier at the output (denoted CCID93), we achieved competitive noise and gain performance at 625 kHz readout speed \cite{chattopadhyay23_sisero,Chattopadhyayetal2022}. We also demonstrated RNDR, achieving a readout noise of 0.51 $\mathrm{e}^{-}_{\mathrm{RMS}}$ with around 1 $\mu$s correlated double sampling (CDS) per cycle, and 57 RNDR cycles at a detector temperature of -100$^\circ$C \cite{chattopadhyay24_rnrdr,chattopadhyay24_rndr_spie}.        
Our second-generation SiSeRO detectors mature this technology, incorporating modified designs \cite{Donlonpie2024} that seek to overcome some limitations seen in the first devices. 
Variants include devices with 16 parallel SiSeRO outputs (CCID93++B) and others with 16 SiSeRO amplifiers in series at a single output (CCID93++C). 
The SiSeRO technology also has the potential to be incorporated into active pixel sensor (APS) arrays, with each pixel having a SiSeRO amplifier, potentially enabling even faster, very low noise performance. Our second-generation devices include the first $8\times8$ pixel SiSeRO APS device (denoted SiSeRO-APS). 
Fabrication of these second-generation SiSeRO devices is now complete and packaging and preliminary testing are underway. The devices should be available for detailed tests and characterization by the end of the year. 

To date, we have been using custom designed discrete readout electronics to test and characterize SiSeRO devices (see Chattopadhyay et al 2022\cite{chattopadhyay22_sisero} for details of the electronics). However, the new multi-output SiSeRO devices require a fast, low power, low noise, small footprint, multi-channel integrated readout system. At Stanford, we have developed an 8-channel application-specific integrated circuit (ASIC) known as the Multi-Channel Readout Chip or MCRC\cite{herrmann20_mcrc,Oreletal2022,porelMCRCspie2024}. Primarily designed for reading out large format multi-channel X-ray CCDs\cite{miller2023_spie_axis,stueber2025_pjfet_mcrc}, the MCRC also has an experimental drain current readout mode intended for use with SiSeRO detectors. As a proof of concept, we here demonstrate the first use of an MCRC chip to read out a first-generation, single output (CCID93) SiSeRO detector, detailing the optimization process and first results. 

In the following sections, we give a brief description of the SiSeRO device working along with the next generation device designs. In Sec. \ref{sec:electronics} and \ref{sec:results}, we discuss the MCRC integrated readout system and the first results from the current SiSeRO detectors respectively. We summarize the work in section \ref{sec:summary}.


\section{An overview of the SiSeRO devices}
\begin{figure}[t!]
    \centering
   \includegraphics[width=.6\linewidth]{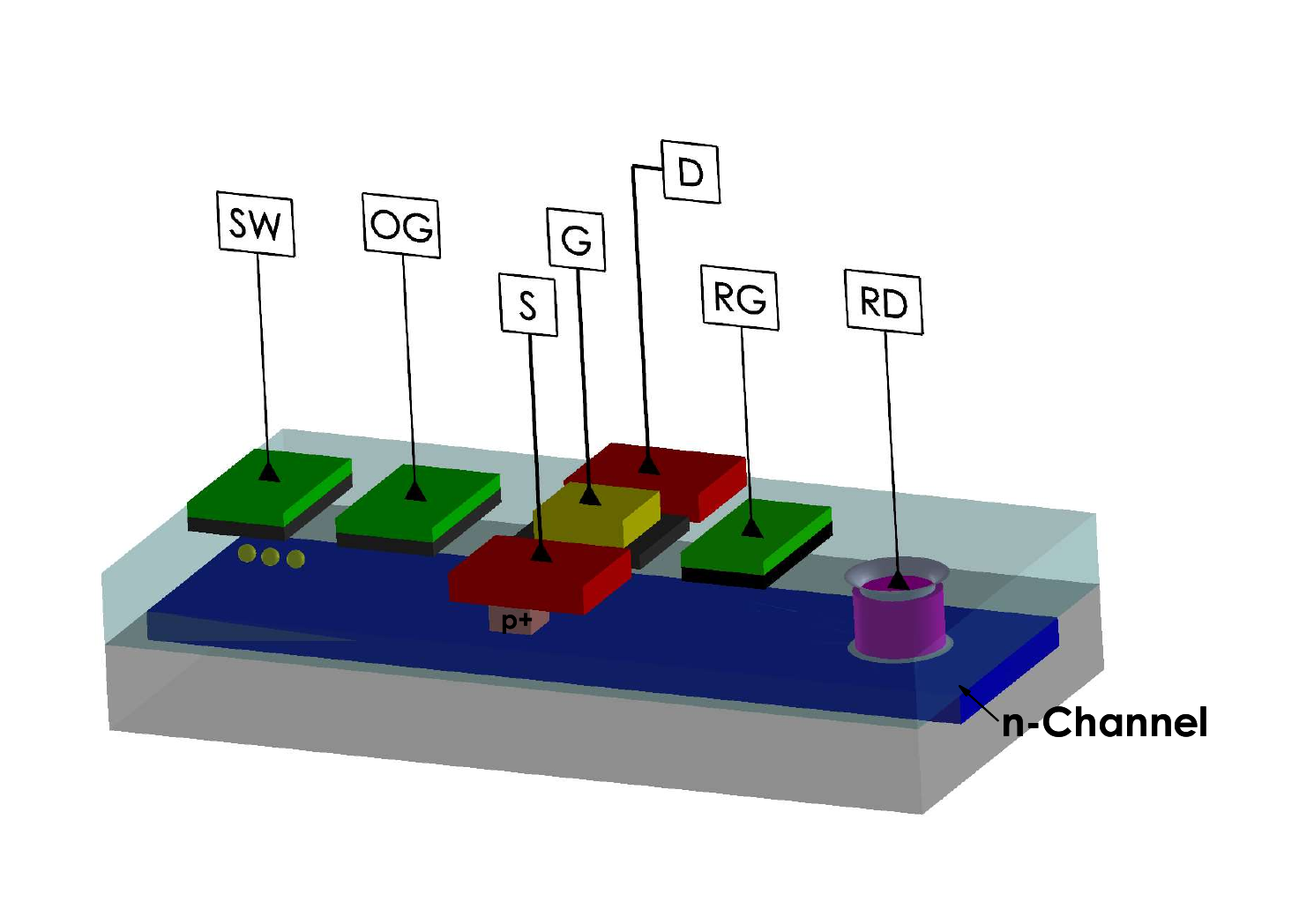}
   \includegraphics[width=.38\linewidth]{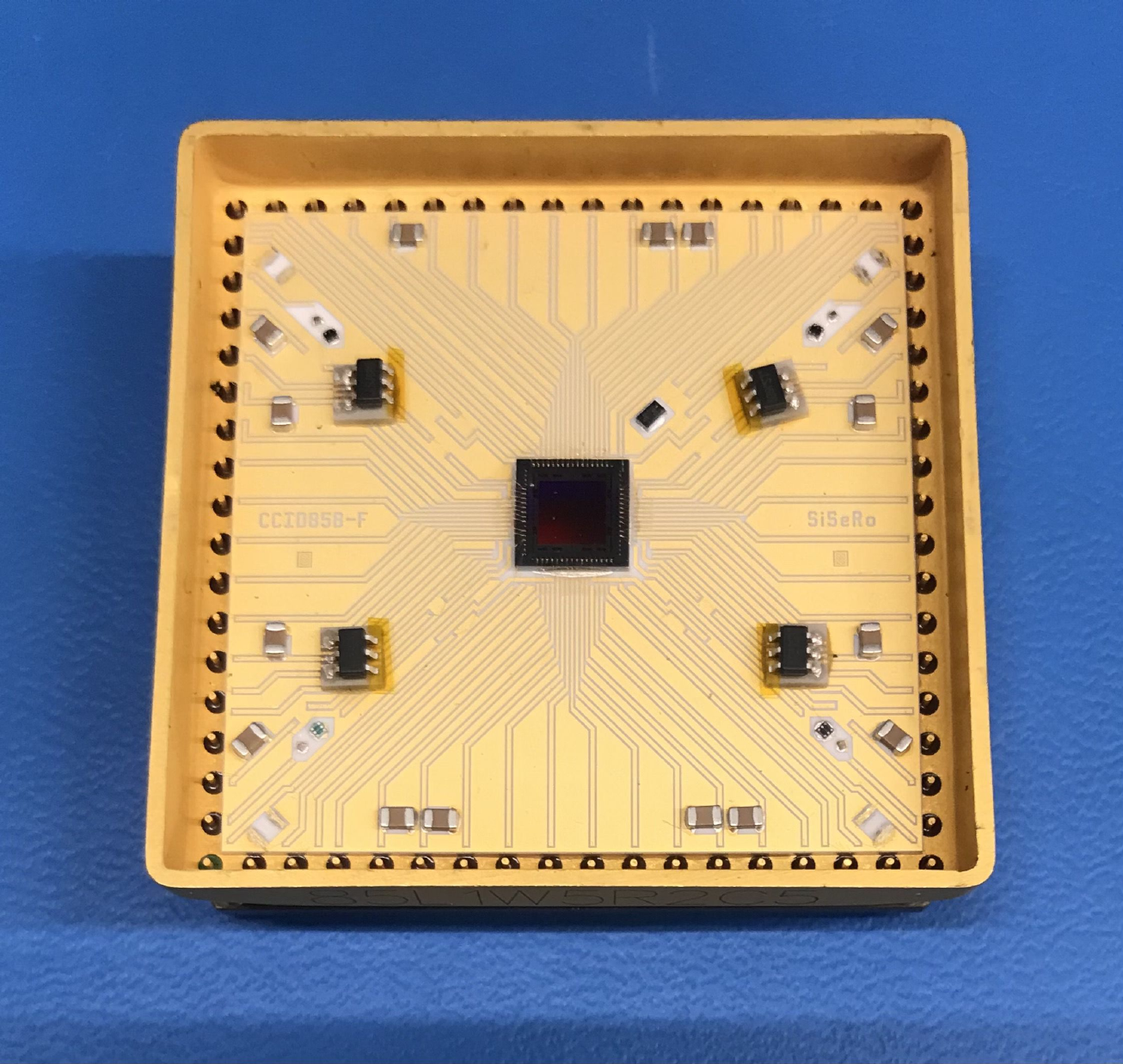}
    \caption{Left: Schematic of a SiSeRO output stage, utilizing a p-MOSFET transistor (S: Source, D: Drain, G: Gate) with an internal gate beneath the p-MOSFET. The internal gate is emptied using the reset transistor (RG) and reset drain (RD). Charge transferred across the output gate (OG) from the last serial gate (summing well or SW) to the internal gate modulates the drain current of the p-MOSFET. Right: A first generation SiSeRO device.}
    \label{fig:sisero}
\end{figure}
The first-generation SiSeRO detectors \cite{chattopadhyay22_sisero} consist of a CCD pixel array with a SiSeRO amplifier at the output stage. SiSeROs use a p-MOSFET transistor with a depleted internal gate under the MOSFET gate. Figure \ref{fig:sisero} (left) shows a cartoon 3D schematic of a SiSeRO output stage. As a charge packet is transferred to the internal gate, it modulates the transistor drain current. Custom off-chip drain current readout electronics are used to sense that change in current. A trough, typically at the center of the internal gate, localizes the charge packet under the gate.
The internal gate minimizes the parasitic capacitance on the sense node compared to traditional CCD sense nodes (comprising a floating diffusion and source follower circuit). This results in a higher conversion gain and improved noise performance. An interesting feature of the SiSeRO technology is that the charge signal remains unaffected during the readout process, allowing the possibility to read the same charge multiple times by moving the charge in and out of the internal gate, enabling repetitive, non-destructive readout (RNDR). By averaging over a series of N independent measurements, the final noise is reduced by a $\sqrt{N}$ factor, at the expense of an increased total readout time.  

Figure \ref{fig:sisero} (right) shows a first generation SiSeRO device, denoted CCID93, fabricated at the MIT-LL. It is an n-channel p-substrate device with a 512 $\times$ 512 array of 8 $\mu$m pixels and one SiSeRO output stage using a buried channel p-MOSFET transistor. For these test devices, we achieved 4.5 $\mathrm{e}^{-}_{\mathrm{RMS}}$ readout noise at 625 KHz readout speed for a detector temperature of -25$^\circ$C using discrete readout electronics \cite{chattopadhyay23_sisero}. We measured a gain or responsivity of 800 pico-ampere (pA) per electron and a full width at half maximum (FWHM) spectral resolution of $\sim$132 eV at 5.9 keV (the Mn K$_\alpha$ line). We also demonstrated the first RNDR measurements with these devices \cite{chattopadhyay24_rnrdr}, more recently achieving sub-electron noise performance (0.51 $\mathrm{e}^{-}_{\mathrm{RMS}}$) for 57 RNDR cycles (with a starting noise of 3.75 $\mathrm{e}^{-}_{\mathrm{RMS}}$) with the help of an improved cooling system (-100$^\circ$C)\cite{chattopadhyay24_rndr_spie}. The net readout speed is around 10 KHz for 57 RNDR cycles. It is to be noted that a SiSeRO variant, developed at MIT-LL using a different fabrication technology (n-MOSFET SiSeRO with hole collecting CCD) was also shown to achieve sub-electron noise  \cite{sofoharo2023_nsisero}, demonstrating the robustness of this new detection technology in achieving low noise performance.


\section{Second generation SiSeRO devices}
The second generation of SiSeRO detectors (CCID93++) follow the same form factor as the CCID93 detectors with 512 $\times$ 512 image and frame store arrays of 8 $\mu$m pixels. There are three detector flavors in the CCID93++ family $-$ CCID93++A (hereafter vA), CCID93++B (hereafter vB), and CCID93++C (hereafter vC) \cite{Donlonpie2024}. The vA flavor is designed to mature and test the p-JFET CCD technology further, whereas the vB and vC flavors are dedicated to SiSeRO technology maturation. Below we discuss the vB and vC SiseRO devices briefly.

\begin{figure}[h!]
\centering
    \includegraphics[width=0.6\textwidth]{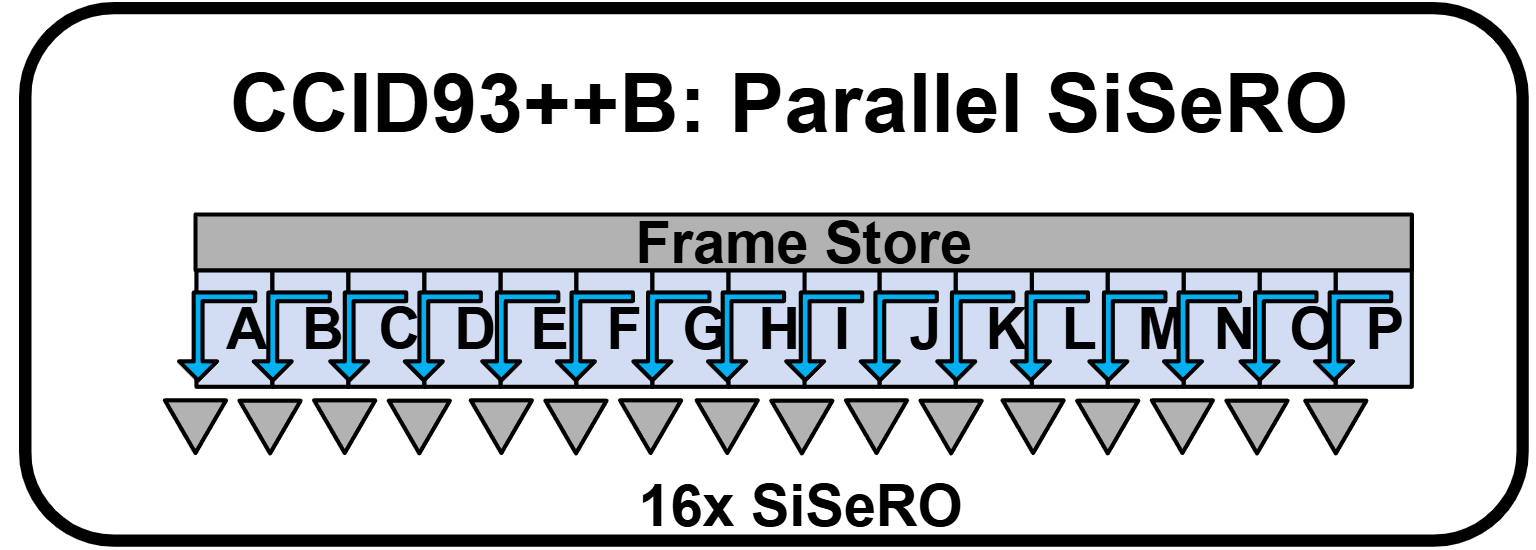}
    \caption{Cartoon rendition of the CCID93++B SiSeRO output stage. Sixteen outputs are grouped into four geometries: A-D, E-H, I-L and M-P (see text for details). The
devices have active imaging areas and frame store areas of 512$\times$512 pixels, with a pixel size of 8 $\mu$m.}
\label{fig:SiSeRO_par}
\end{figure}

In vB, there are 16 total outputs designed to run in parallel, with each output having a SiSeRO amplifier. The 16 outputs are built in four sets of four
SiSeRO variants (see Fig. \ref{fig:SiSeRO_par}), 
where the pitch between the output stages is around 256 $\mu$m (512 serial pixels divided by 16 output nodes). 
In the first geometry, the design is the same as the first-generation CCID93 SiSeRO legacy devices. The other geometrical variants explore: (1) a different location for the charge packet on the internal gate (0.5 $\mu$m away from the center towards the source) in order to enhance the conversion gain; (2) 
enhanced isolation of the p-MOSFET source/drain and the p-type channel stop region to minimize the leakage paths yielding parasitic transistor current (see Fig. 5 of Chattopadhyay et al. 2022 \cite{chattopadhyay22_sisero}); and (3) reduction in the p-MOSFET
gate resistance, leading to faster response of gate potential on the channel current. 
The internal gate in all of these amplifier geometries is located deeper into the substrate (away from the silicon surface interface) to minimize the 1/f noise contribution from trapping and de-trapping of charge carriers (see Fig. 4 in Chattopadhyay et al 2023\cite{chattopadhyay23_sisero}). 

\begin{figure}[h!]
\centering
    \includegraphics[width=0.49\textwidth]{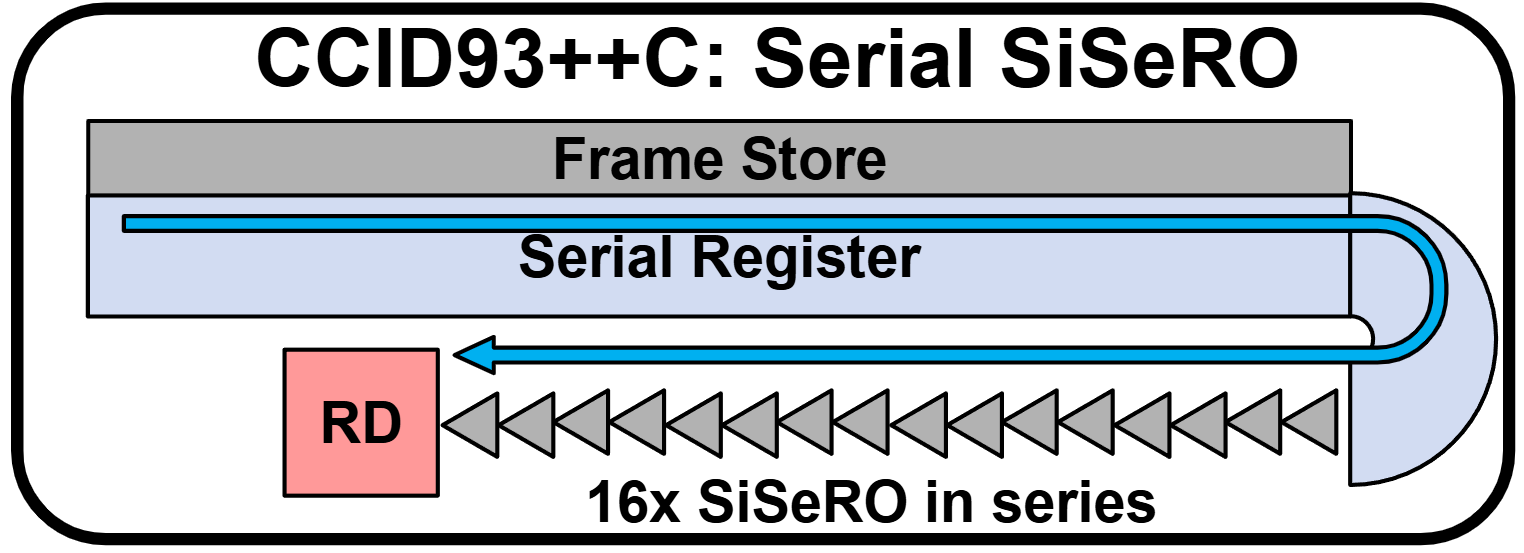}
    \includegraphics[width=0.49\textwidth]{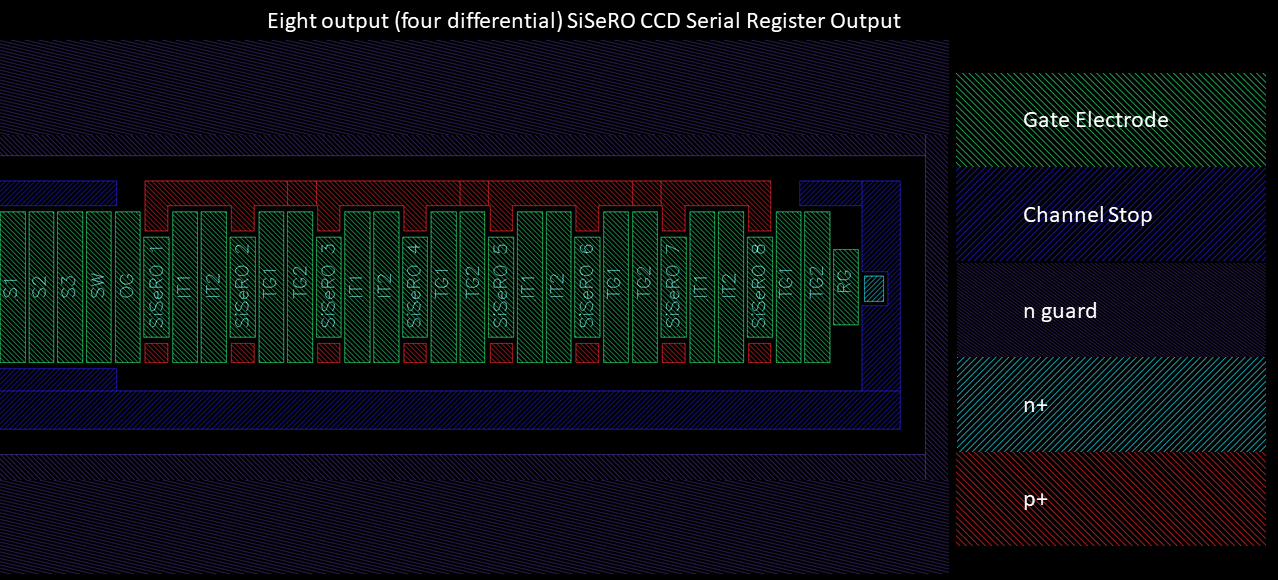}
    \caption{Left: Cartoon rendition of the CCID93++C SiSeRO output stage. Here the 16 output stages are identical. As a charge packet passes through the output nodes, 16 independent measurements of the same charge are made. This allows for 16 RNDR cycles of the charge at the full readout speed. Right: A layout of the vC device showing only eight outputs. Same as vB, these devices have an active imaging area and frame store area of 512$\times$512 pixels with pixel size of 8 $\mu$m.}
\label{fig:SiSeRO_ser}
\end{figure}

The vC devices have 16 SiSeRO amplifiers (of a single variant) in series, deliberately targeting the RNDR capability of
SiSeRO amplifiers. A cartoon rendition of the design is shown in Fig. \ref{fig:SiSeRO_ser} (left). 
In this design, a charge packet is passed through all the  amplifier stages non-destructively providing sixteen independent measurements of the same charge. 
As the second amplifier in the chain probes the charge packet, the next pixel's charge packet is received by the first amplifier simultaneously and passed on through the chain. 
This gives a reduction in read noise by a factor $\sqrt{16}=4$; which results in a read noise of $\sim$3.75/4 $\sim 1$ $\mathrm{e}^{-}_{\mathrm{RMS}}$ at 625 KHz. As shown in the layout of the output stage in Fig. \ref{fig:SiSeRO_ser} (right), there are dedicated transfer gates between the amplifiers to facilitate the transfer of charge as well as repetitive readout of the charge by the same amplifier multiple times as in case of a traditional single amplifier RNDR device. This allows for further reduction in noise, although at the expense of readout speed. The pitch between two adjacent amplifiers is 27 $\mu$m.

\begin{figure}[h!]
\centering
    \includegraphics[width=0.48\textwidth]{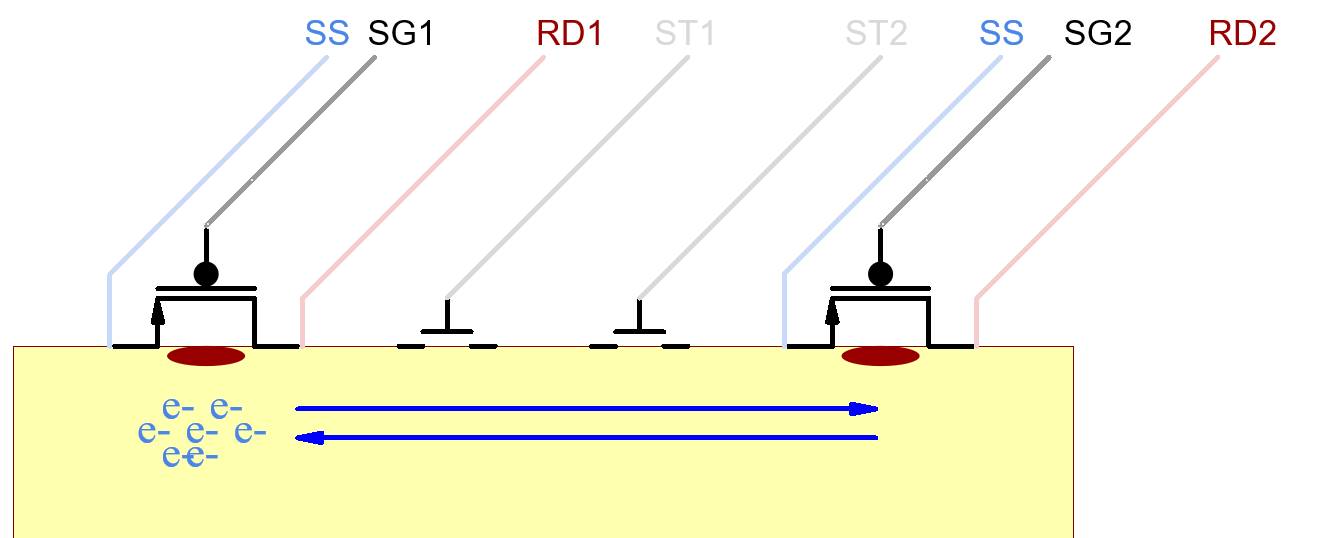}
    \includegraphics[width=0.5\textwidth]{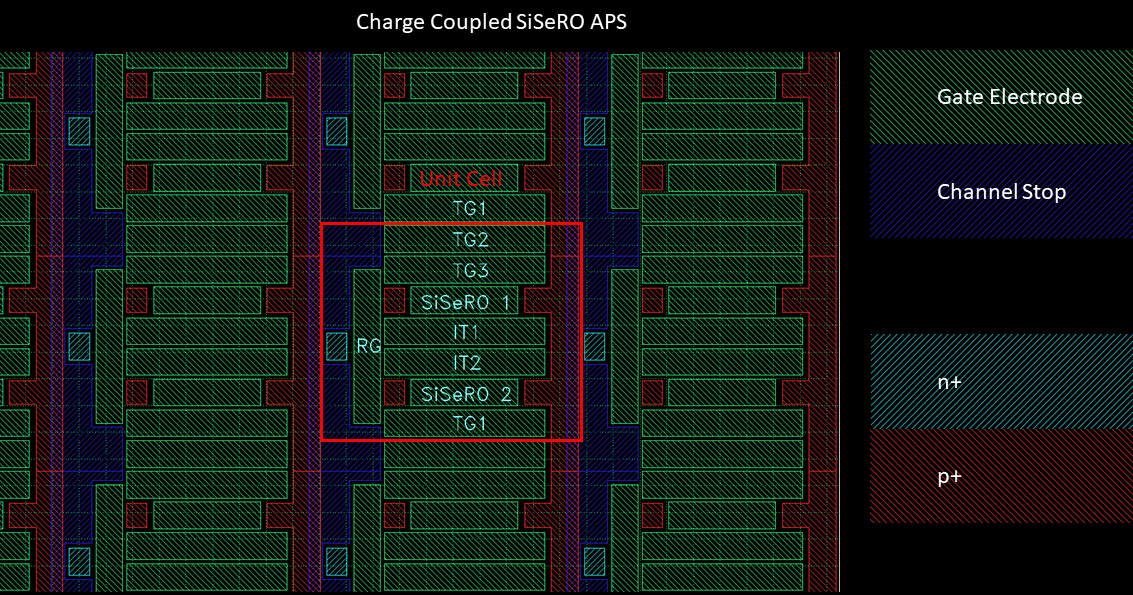}
    \caption{RNDR optimized SiSeRO APS. Left: Schematic of two p-MOSFET SiSeRO transistors placed next to each other, with a dedicated transfer gate arrangement between them, allowing signal charge to be moved between the two amplifiers for repetitive measurements. Right: Layout sketch of with two SiSeROs at every pixel (denoted by the red square box). The pixel boundaries are established by a channel stop (blue) and barrier gate structure (TG1).}
\label{fig:SiSeRO_matrix}
\end{figure}
We have also designed our first SiSeRO active pixel sensor (APS) following the concept used in RNDR optimized DEPFET detectors \cite{wolfel06}. SiSeRO APS with RNDR provides the capability to read out larger detector arrays with low-noise at significantly higher speed, along with region-of-interest readout in the same observation. Such a device with no macro charge transfer is also radiation hard and largely immune to radiation damage induced charge transfer inefficiency.
In our APS design, each pixel has two SiSeRO amplifiers next to each other, allowing RNDR by transferring the charge between the two transistors.  A cartoon schematic of the concept is shown in left side of Fig. \ref{fig:SiSeRO_matrix}, whereas the right side shows a layout for the structure. 
There are two internal transfer gates (IT1, IT2) between the transistors to allow efficient charge transfer. Since the SiSeROs are immune to reset noise \cite{chattopadhyay23_sisero}, the signal charge and baseline can be measured simultaneously from the two transistors respectively, allowing for faster RNDR switch frequencies (by a factor of two). The first SiSeRO APS is 8 $\times$ 8 array in size with 24 $\mu$m pixel size, currently under fabrication in the same CCID93++ fabrication cycle. The pixel separation on the sides is achieved with the column parallel channel stops, whereas a barrier gate (TG1) provides separation on top and bottom. Each pixel has a reset gate and drain on the side.    

Fabrication of the first devices is already complete and currently being packaged. We expect the devices to be ready for testing by the end of the year. These devices are currently front illuminated. However, note that the fabrication is compatible with MIT-LL's molecular beam epitaxy (MBE) back illumination process.


\section{Integrated readout electronics for the second generation SiSeRO devices} \label{sec:electronics}
 To enable parallel readout in the second-generation SiSeRO detectors (vB, vC, and SiSeRO-APS) without increasing the power consumption or area, it is essential to switch from discrete electronics to an ASIC based integrated readout approach. The Stanford MCRC chip\cite{herrmann20_mcrc,Oreletal2022} offers eight drain readout channels. Consequently, two MCRC chips can be used to read out the sixteen SiSeRO nodes in the vB and vC devices. The readout of the SiSeRO-APS will require a novel integrated readout solution. 
To prove the concept of reading out a SiSeRO amplifier with an MCRC chip, at Stanford, we have developed a readout electronics board that offers simple switching between a readout circuit comprising discrete components and an integrated circuit using the MCRC chip. Each channel of the MCRC chip has two programmatically selectable inputs: a source follower (SF) input (for traditional voltage-based, high impedance CCD readout) and an experimental drain current (DR) input for SiSeRO readout. The theory of operation of the SF input and its optimization have been discussed in detail in Porel et al 2024\cite{porelMCRCspie2024}. Here we briefly discuss the working of the DR input mode. 

\begin{figure} [ht]
   \begin{center}
   \includegraphics[width=0.7\textwidth]{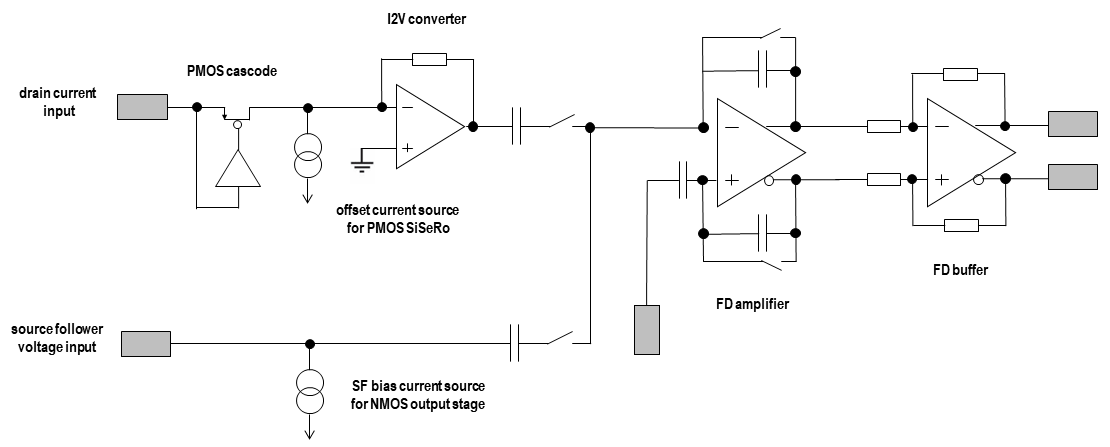}
   \includegraphics[width=0.45\textwidth]{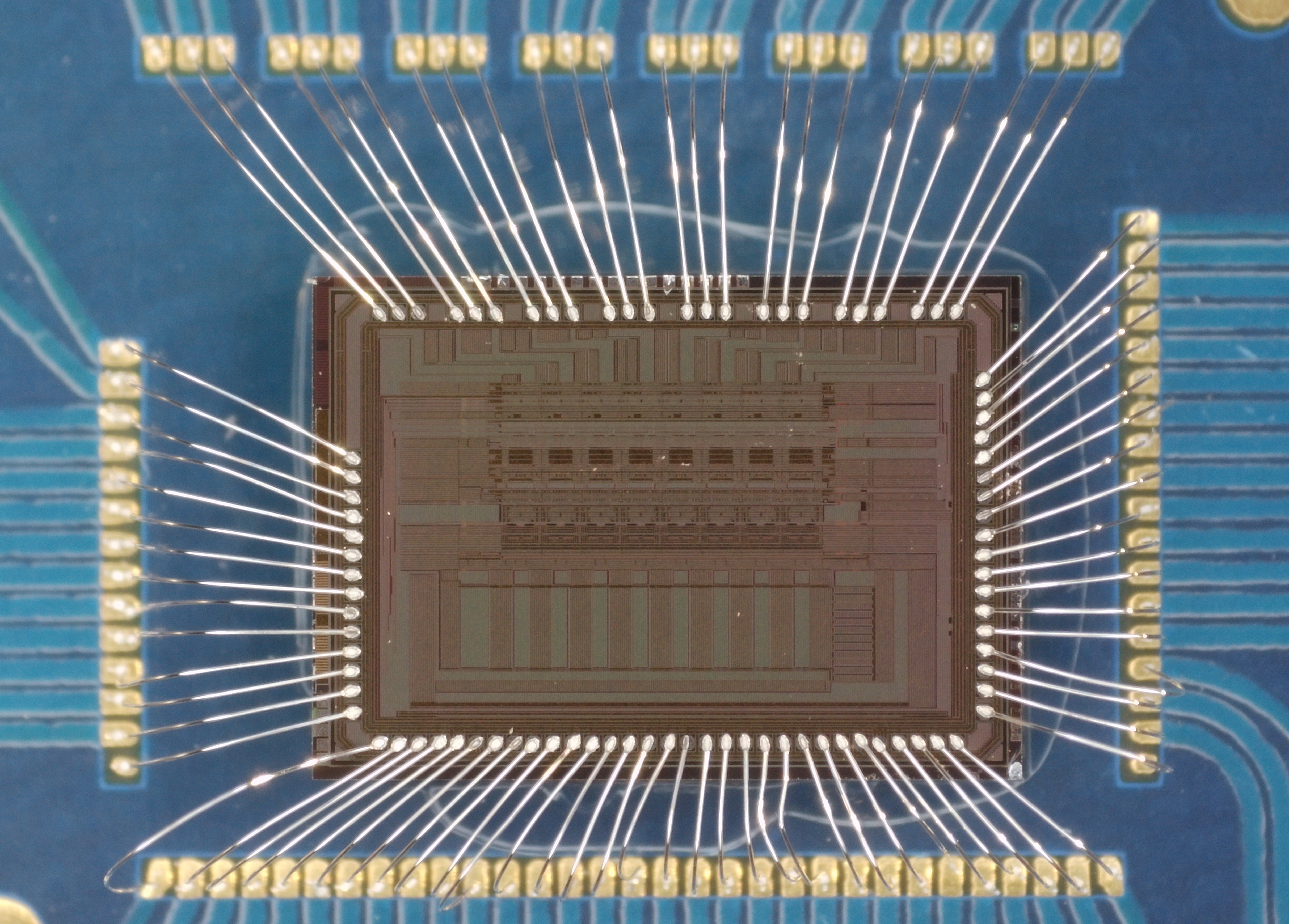}
   \end{center}
   \caption{Top: The {MCRC-V1} single analog channel schematic \cite{herrmann20_mcrc}. Bottom: Photograph of the fabricated {MCRC-V1} die. The inputs are located on the top side of the chip, while the outputs are on the bottom. The left side has the bias and current source pads, while the SPI interface is on the right side.}
   \label{fig:mcrcsch}
   \end{figure}

Figure~\ref{fig:mcrcsch} (top) shows the schematic of a single channel.
The DR input features an active cascode (AC) that isolates the sensitive current sensing node of the following current to voltage (I2V) amplifier, as well as providing constant drain voltage to the SiSeRO drain. The AC is followed by a current source designed to provide the necessary bias current to the keep the SiSeRO output at the ideal operating point. The I2V converts the SiSeRO current signal into a voltage signal, with a conversion gain (transresistance) of 100 k$\Omega$. The I2V output signal is then ac-coupled to a fully differential pre-amplifier (PRE) circuit, which converts the signal from single-ended to differential. The PRE has two user selectable gain settings: 8 V/V (G8) and 16 V/V (G16). The PRE has a switch capacitor based feedback and needs to be reset at regular intervals to avoid saturation. The differential signal is buffered to the outside by a unity gain fully differential output buffer (OB), which is designed to drive a 100 $\Omega$ differential transmission line and interface directly to an analog to digital converter (ADC). In the first-generation MCRC-V1 chip, a p-MOSFET based AC circuit provides a constant bias of $\sim$1 V, which is independent of the SiSeRO current. To bias the drain at to a different voltage level, the entire MCRC ground reference can be shifted to accommodate the appropriate bias level. All the MCRC signal chain circuits (AC, I2V, PRE, and OB) can be biased independently using four programmable internal DACs. The digital core of the chip uses a serial peripheral interface (SPI) bus to communicate and configure the ASIC.              

Figure \ref{fig:mcrcsch} (bottom) shows a photograph of the fabricated MCRC-V1 chip. The chip is 4160 $\mu$m by 2900 $\mu$m in size, designed and manufactured in the X-FAB ultra-low-noise CMOS XH035 3.3 V 350 nm process technology\cite{xfab}.
The analog core of the chip is at the center with eight inputs on the top side and the corresponding outputs on the bottom side of the chip. The left and the right sides are dedicated for the bias / current source pads and the digital interface. 

To improve modularity, redundancy, and ease of use, the chip is not directly wire bonded to the detector readout board. Instead it is wire bonded to a separate carrier board, which is then mounted on the readout board using a low-profile interposer\footnote{\url{https://www.samtec.com/products/gmi}}. This provides flexibility to keep the MCRC and readout board separate, allowing for the removal or replacement of the chip if needed without the need to re-fabricate the detector board. The detector board can be seen in Fig. \ref{fig:setup} (right) inside the vacuum chamber, hosting the MCRC carrier board and the detector. On the board, we separate the MCRC and detector grounds and connect the MCRC ground to an adjustable DC bias to optimize the SiSeRO drain bias voltage. A commercial controller (Archon controller\cite{archon14}), the black box shown in the left picture, procured from Semiconductor Technology Associates, Inc (STA\footnote{\url{http://www.sta-inc.net/archon/}}), is used to provide the required bias voltages, clock signals, and the digitization of the analog output signals. A custom designed flex cable is used to interface the readout board to the controller.

\section{Proof-of-concept results from the integrated readout system}
\label{sec:results}

\begin{figure}[t]
    \centering
    \includegraphics[width=0.68\linewidth]{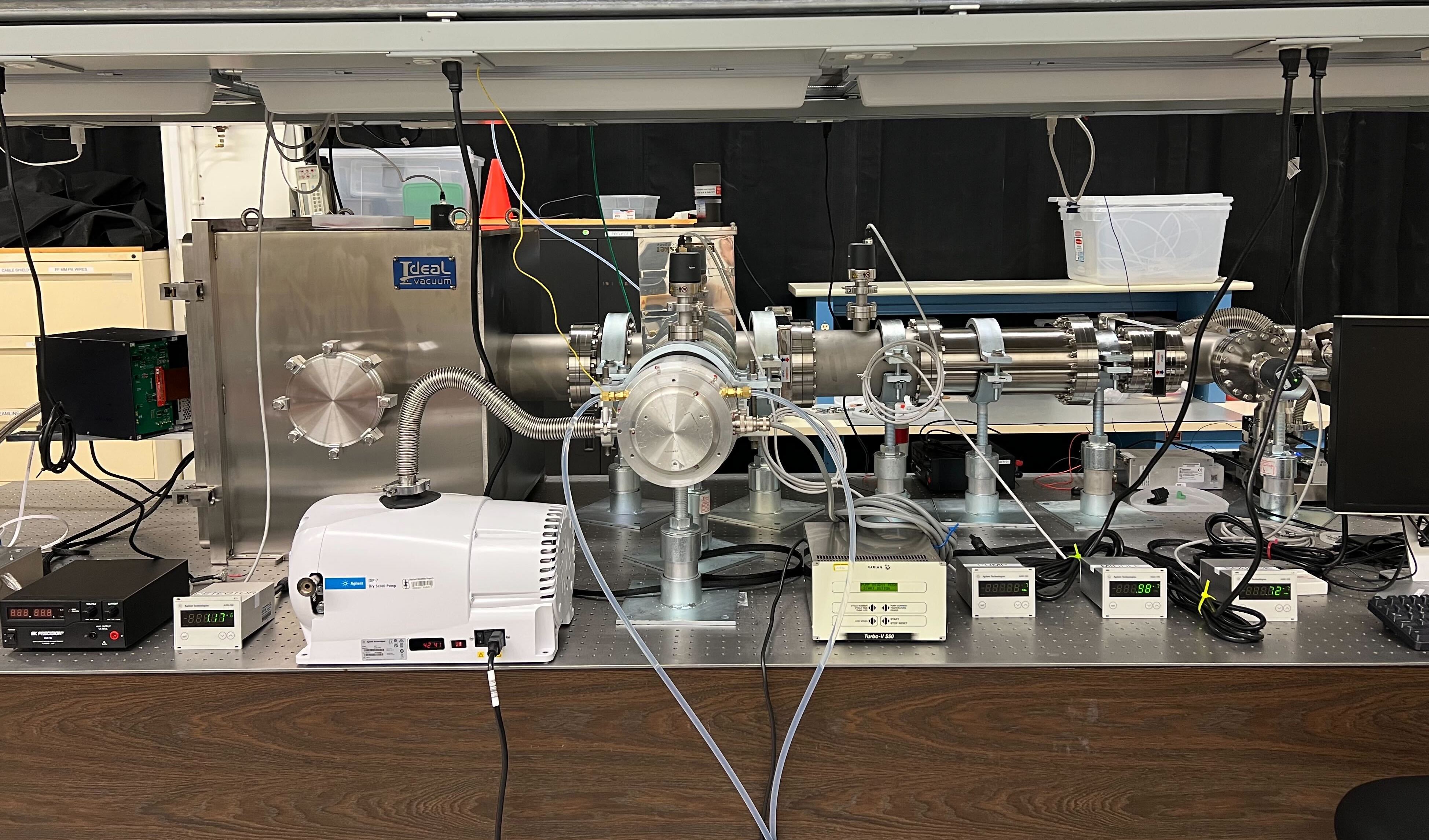}
    \includegraphics[width=0.3
    \linewidth]{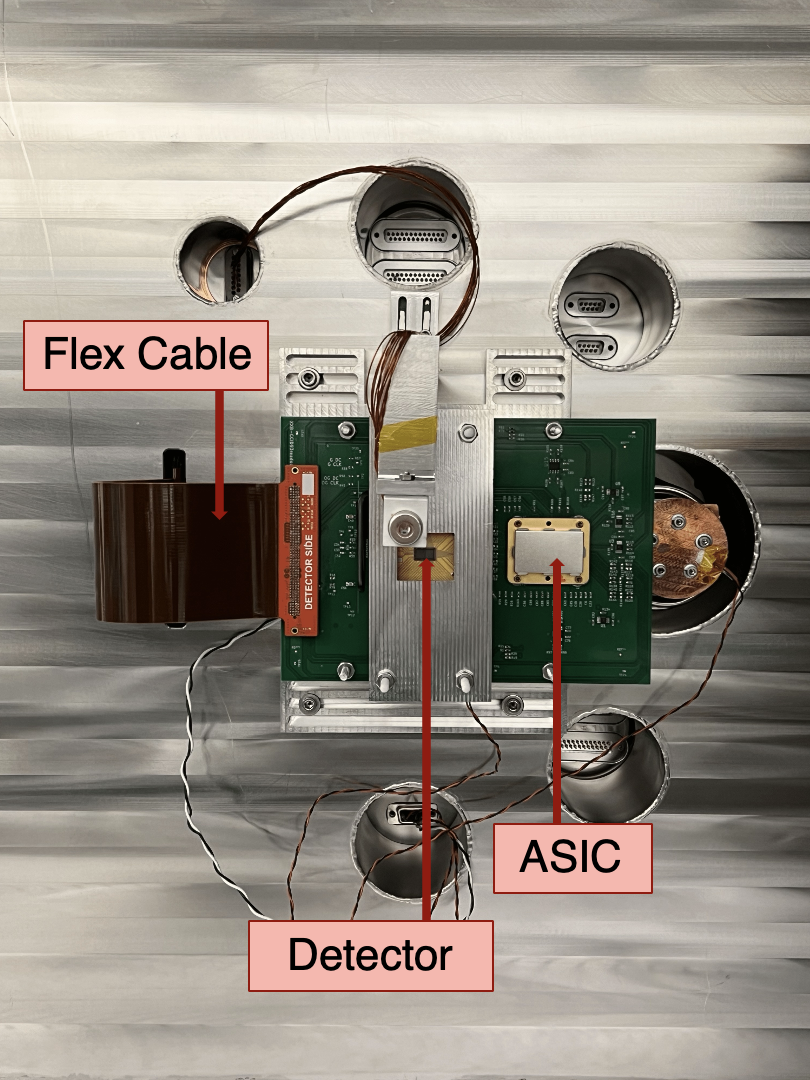}
    \caption{Left: The 2.5 meter XOC X-ray Beamline\cite{stueber2024} used for the tests. Right: A CCID-93 single output SiSeRO device and the MCRC chip integrated readout mounted inside the detector chamber on the chamber door.}
    \label{fig:setup}
\end{figure}

Our measurements of the first-generation SiSeRO plus MCRC ASIC system were made using   
the 2.5 meter XOC X-ray Beamline at Stanford \cite{stueber2024} (Fig. \ref{fig:setup}). 
The detectors were cooled to -100$^\circ$C using a compact cryo-cooler system. A Proportional-Integral-Derivative (PID) regulator keeps the detector temperature stable with an accuracy better than 0.2$^\circ$C. For X-ray measurements, we used an $^{55}$Fe radioactive source mounted inside the chamber. 
 
For these tests, the readout speed of the detector was set to 625 KHz.
We biased the current source of the MCRC (see Sec. \ref{sec:electronics}) at -12.5 V resulting in a current of 62.5 $\mu$A. The MCRC was configured with an AC bias current of 56 $\mu$A, I2V bias current of 83 $\mu$A, PRE bias current of 18 $\mu$A, and OB bias current of 29 $\mu$A. The gain of the PRE was set at 8 V/V and is reset at the end of every line of the detector. The SiSeRO p-MOSFET source and gate were biased at 4 V and 3 V, respectively. The drain was biased at $\sim$1 V through the AC circuit.  
\begin{figure}[t]
    \centering
 \includegraphics[width=.47\linewidth]{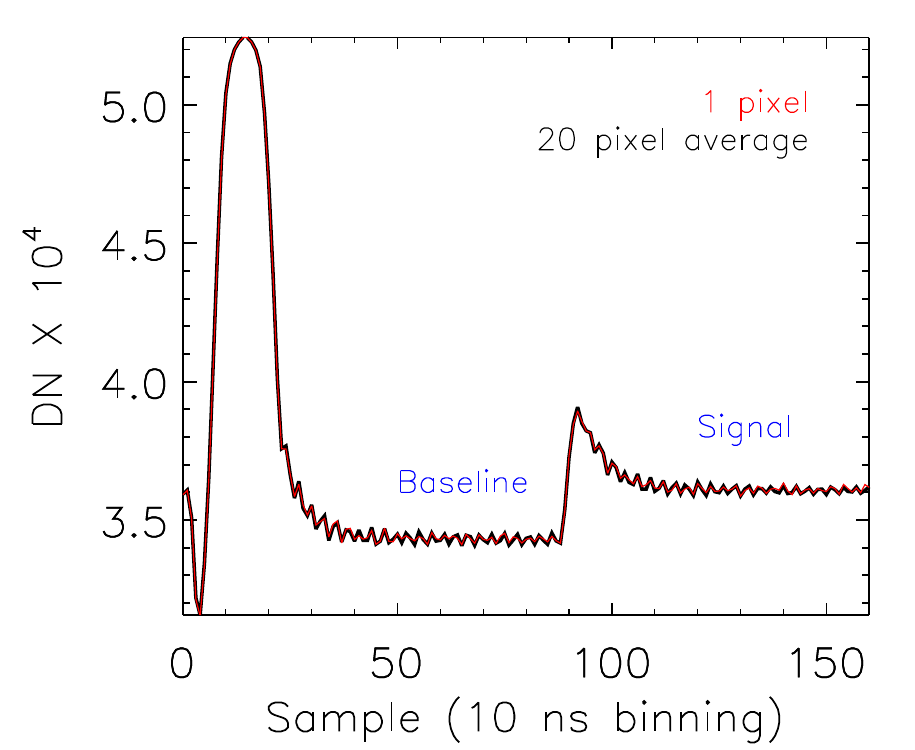}
 \includegraphics[width=.515\linewidth]{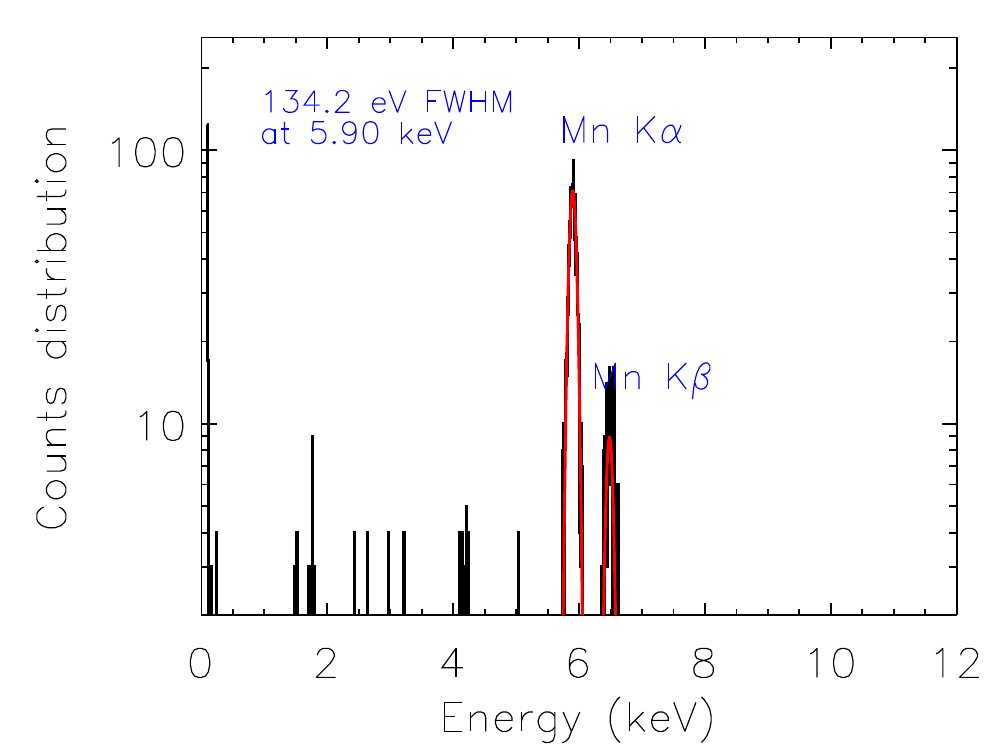}
    \caption{Left: Digital waveform of the detector at 625 KHz readout speed. The red and black lines represent one pixel and twenty pixel averaged waveforms respectively. The oscillations seen in the waveform are in phase. Right: Spectrum obtained at -100$^\circ$C showing the Mn K$_\alpha$ (5.9 keV) and K$_\beta$ (6.4 keV) lines. The spectra were generated using the single-pixel (grade 0) events. The FWHM energy resolution at 5.9 keV is 134 eV and the corresponding read noise is 3.5 $\mathrm{e}^{-}_{\mathrm{RMS}}$.}
    \label{fig:spectra_625}
\end{figure}
Figure \ref{fig:spectra_625} (left) shows digital waveform of the SiSeRO obtained from the Archon controller at 625 KHz at -100$^\circ$C. The ADC is sampled at 100 MHz (10 ns sampling). The red line is the waveform for one single pixel while the black line is obtained when averaged over twenty pixels. At the start of the pixel, the sense node is reset to a high voltage (RD) for a short time interval, immediately after which the internal gate settles down to the baseline region. As we transfer the charge packet from the last serial gate (SW) to the internal gate, the waveform settles to the signal region. The charge signal is measured by digitally sampling the two regions with equal weights and taking a difference between them (this process is popularly known as correlated double sampling or CDS). 

Despite having the active cascode, the capacitive loading on the I2V input proved to be still be an issue. To shield the I2V further, we added a 4 k$\Omega$ resistor at the very close to the input of the I2V. This helped to minimize oscillations initially seen in the detector output signal. The residual high frequency oscillations (35 $-$ 40 MHz) seen in the waveform are attributed to a resonance that is present in the I2V frequency response, which will be resolved in the next MCRC version. These high frequency oscillations are outside of the bandwidth of interest and thus do not contribute significantly to the total added noise of the system. It has to be noted that a voltage drop $\sim$ 0.3 V (4 k$\Omega~\times$~80 $\mu$A, where 80 $\mu$A is the SiSeRO current) on the 4 k$\Omega$ resistor results in a change in the drain bias voltage to $\sim$1.3 V. For this bias setting, we measure the read out noise of the detector to be $\sim3.5$ $\mathrm{e}^{-}_{\mathrm{RMS}}$. 

Figure \ref{fig:spectra_625} (right) shows single-pixel (grade 0) spectra of the Mn K$_\alpha$ and K$_\beta$ lines at 5.9 and 6.4 keV, respectively. The Mn K$_\alpha$ was fit with a Gaussian model to calculate the centroid and FWHM of the line, which was found to be 134 eV.  

In Table \ref{table:RO_compare}, we compare the results obtained from the integrated MCRC and discrete readout systems for a 625 kHz readout speed and -100$^\circ$C detector operating temperature.  
\begin{table}[]
\centering
\caption{Comparison of the results obtained using the MCRC and discrete readout systems, for a 625 KHz readout speed and detector temperature of -100$^\circ$C. Here ``GND Offset" refers to the offset between the MCRC and detector grounds. V$_S$ and V$_D$ stand for SiSeRO source and drain bias voltages respectively. ``ADU" is the digital unit of the charge. The measured error on the read noise is typically around 2 \% and around 1-2 \% on the FWHM.}
\label{table:RO_compare}
\begin{tabular}{|l|cccc|c|}
\hline
             & \multicolumn{4}{c|}{MCRC Readout}                                                                                                                                                                                                                                                                                                                                                                                               & Discrete Readout                                               \\ \hline
             & \multicolumn{1}{c|}{\begin{tabular}[c]{@{}c@{}}GND Merged\\ (V$_S$=4V,\\ V$_D$$\sim$1.3V\end{tabular}} & \multicolumn{1}{c|}{\begin{tabular}[c]{@{}c@{}}GND offset \\ by 0V\\ (V$_S$=4V, \\ V$_D$$\sim$1.3V)\end{tabular}} & \multicolumn{1}{c|}{\begin{tabular}[c]{@{}c@{}}GND offset \\ by -0.5 V\\ (V$_S$=4V, \\ V$_D$$\sim$0.8V)\end{tabular}} & \begin{tabular}[c]{@{}c@{}}GND offset \\ by +0.5 V\\ (V$_S$=4.5V, \\ V$_D$$\sim$1.8V)\end{tabular} & \begin{tabular}[c]{@{}c@{}}(V$_S$=4V, \\ V$_D$=1V)\end{tabular} \\ \hline
Read Noise   & \multicolumn{1}{c|}{3.53 $\mathrm{e}^{-}_{\mathrm{RMS}}$}                                        & \multicolumn{1}{c|}{3.54 $\mathrm{e}^{-}_{\mathrm{RMS}}$}                                                   & \multicolumn{1}{c|}{3.68 $\mathrm{e}^{-}_{\mathrm{RMS}}$}                                                       & 4.65 $\mathrm{e}^{-}_{\mathrm{RMS}}$                                                         & 3.54 $\mathrm{e}^{-}_{\mathrm{RMS}}$                           \\ \hline
5.9 keV Gain & \multicolumn{1}{c|}{5190 adu}                                                                    & \multicolumn{1}{c|}{5227 adu}                                                                               & \multicolumn{1}{c|}{5662 adu}                                                                                   & 4019 adu                                                                                     & 2069 adu                                                       \\ \hline
5.9 keV FWHM & \multicolumn{1}{c|}{134 eV}                                                                      & \multicolumn{1}{c|}{136 eV}                                                                                 & \multicolumn{1}{c|}{139 eV}                                                                                     & 141 eV                                                                                       & 134 eV                                                         \\ \hline
\end{tabular}
\end{table}  
\begin{figure}[h!]
    \centering
 \includegraphics[width=.7\linewidth]{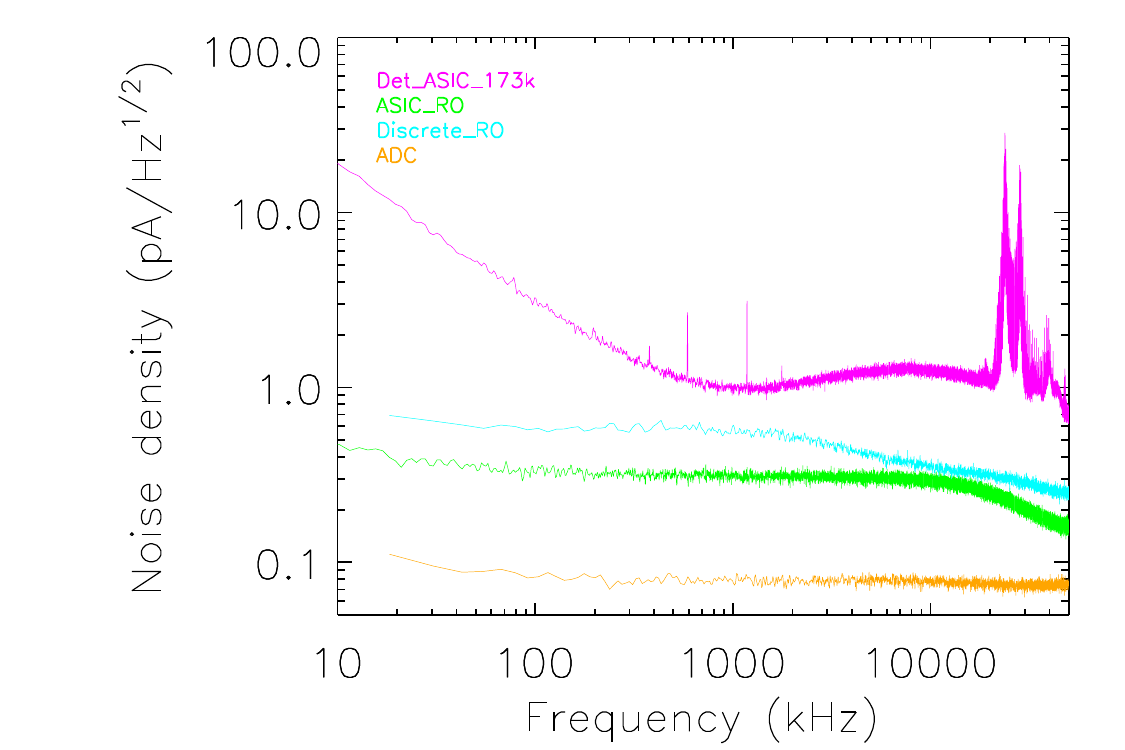}
    \caption{Power spectrum density of the detector and the readout electronics. The total noise is dominated by the detector noise; the ASIC noise is a magnitude lower. The peaks at higher frequencies ($>$20 MHz) are probably the oscillations we see in the detector signal waveform. }
    \label{fig:PSD}
\end{figure}
Measurements were performed for four different configurations of the ASIC by shifting the MCRC ground reference to adjust the SiSeRO drain bias voltage (V$_{D}$). 
The last column represents the results obtained from the discrete readout mode. For similar detector biasing conditions, we see the noise and spectral performance of the detector are consistent in both the integrated and discrete readouts. Note that the electronic gain of the discrete system is around a factor of two lower than the MCRC system, which explains the difference in the measured 5.9 keV gains from the two readout modes. 

In Fig. \ref{fig:PSD}, we show the power spectrum density of the detector in the MCRC readout mode (magenta). We also compare the PSDs of the MCRC (green) and discrete (cyan) modes without the detector. The ADC only noise is shown in yellow.    
We achieve comparable or slightly better noise performance using the MCRC readout (green). However, since the total noise is largely dominated by the detector output noise, we get comparable noise performance from the detector for both the MCRC and discrete readout. The high frequency peaks on the far end of the noise spectra are believed to originate from the MCRC I2V (see also Fig. \ref{fig:spectra_625} left).  

\begin{figure}[t]
    \centering
 \includegraphics[width=.47\linewidth]{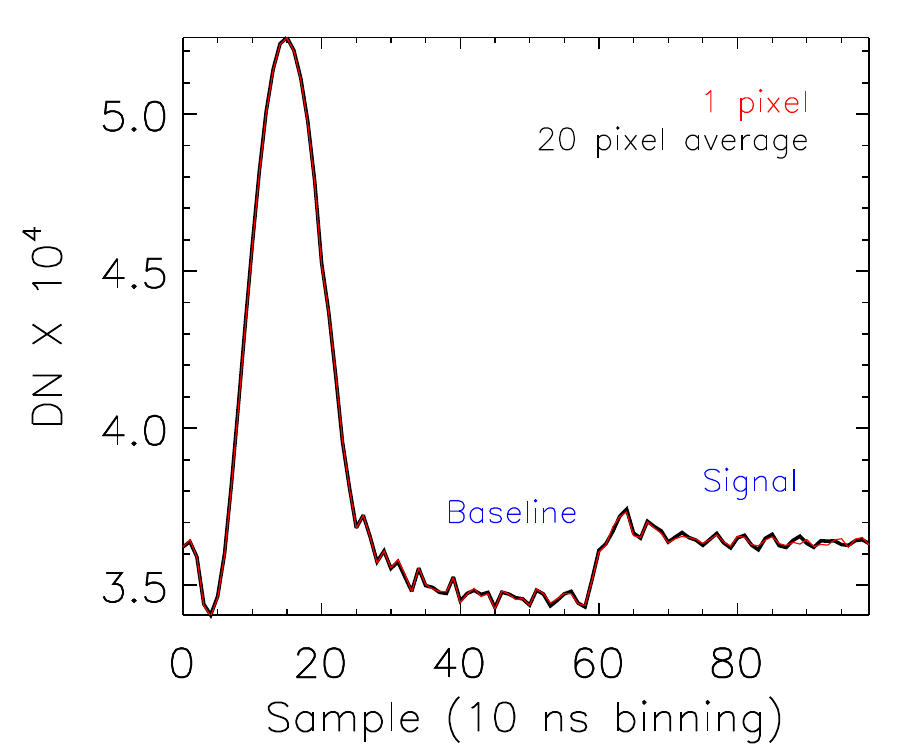}
 \includegraphics[width=.515\linewidth]{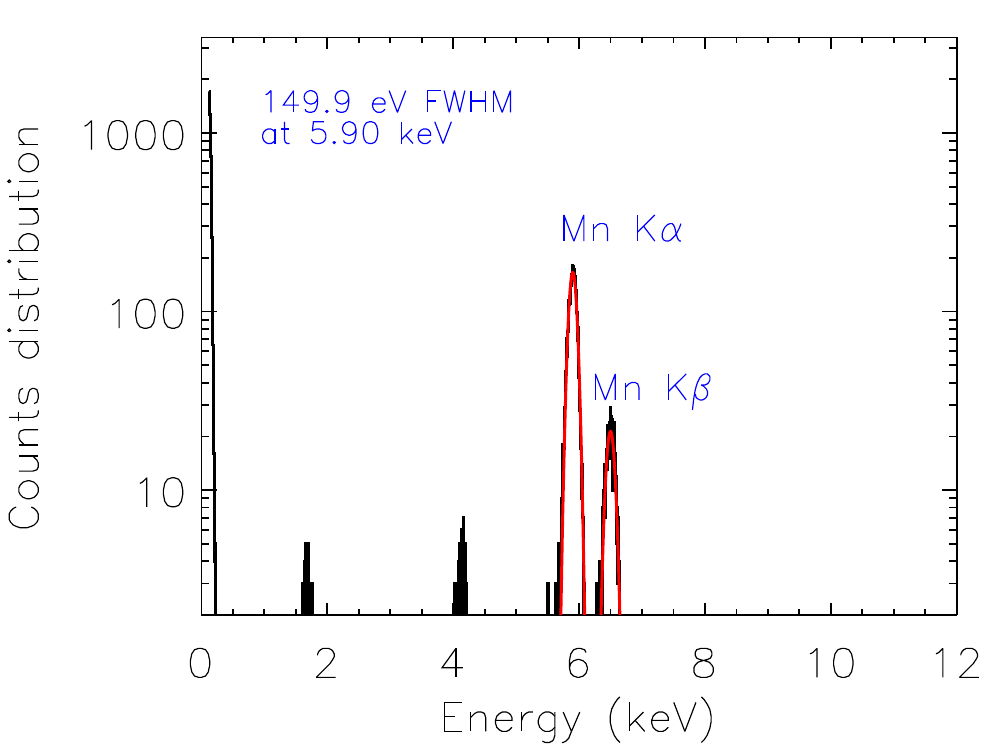}
    \caption{Same as Fig. \ref{fig:spectra_625} but for 1 MHz of readout speed. The FWHM energy resolution at 5.9 keV is 149 eV and the corresponding read noise is around 4.9 $\mathrm{e}^{-}_{\mathrm{RMS}}$.}
    \label{fig:spectra_1mega}
\end{figure}

The MCRC chip offers a bandwidth that should allow us to operate the SiSeROs at higher readout speeds than has been the case using discrete readout electronics. As a proof of concept, for the first time, we ran the detector at 1 MHz readout speed. The MCRC and detector grounds were offset by 0 V for this test. Figure \ref{fig:spectra_1mega} shows the results with the detector at -100$^\circ$C.  
The left plot shows the waveform of a pixel (red) and an average of twenty pixels (black). The tests yielded a noise performance of 4.9 $\mathrm{e}^{-}_{\mathrm{RMS}}$. The single-pixel (grade 0) spectra of the Mn K$_\alpha$ and K$_\beta$ lines are shown in the right plot. The measured FWHM at 5.9 keV is around 149 eV. These results demonstrate the capability of the SiSeRO devices and MCRC to operate at these high readout speeds.   

\section{Summary}\label{sec:summary}
Single electron Sensitive Read Out (SiSeRO) is a novel on-chip charge detection technology offering improved noise and spectral performance compared to traditional CCD readout circuitry. The current first generation SiSeRO devices, developed at MIT-LL, use a CCD pixel array with a single SiSeRO amplifier at the output stage. To mature the technology, we are fabricating new SiSeRO detectors with improved amplifier designs \cite{Donlonpie2024}. Variants include devices with multiple SiSeRO amplifiers at the output to enable rapid RNDR measurements.
We are also building the first SiSeRO APS prototype (8$\times$8 pixel) matrix. Reading out such devices requires low power, low noise, fast, small footprint, multi-channel integrated readout systems. In this paper, we demonstrate such operation using the eight-channel MCRC ASIC chip, developed at Stanford with a first-generation SiSeRO detector. We obtained noise and spectral performance consistent with that of our best discrete readout systems. 

\acknowledgments 
This work has been supported by NASA grants APRA 80NSSC22K1921	``X-Ray Speed-Reading: Integrated Readout Technology for Fast, Very Low-Noise, Megapixel X-Ray Imaging Detectors” and SAT 80NSSC23K0211 ``Extremely Low-noise, High Frame-rate X-ray Image Sensors for Strategic Astrophysics Missions.”




\end{document}